# Numerical Magnetohydrodynamics in Astrophysics: Algorithm and Tests for Multi-Dimensional Flow[1]

Dongsu Ryu

Department of Astronomy and Space Science, Chungnam National University,
Daejeon 305-764, Korea

T. W. Jones and Adam Frank

School of Physics and Astronomy, University of Minnesota, Minneapolis, MN 55455

## ABSTRACT

We present for astrophysical use a multi-dimensional numerical code to solve the equations for ideal magnetohydrodynamics (MHD). It is based on an explicit finite difference method on an Eulerian grid, called the Total Variation Diminishing (TVD) scheme, which is a second-order-accurate extension of the Roe-type upwind scheme. Multiple spatial dimensions are treated through a Strang-type operator splitting. The constraint of a divergence-free field is enforced exactly by calculating a correction via a gauge transformation in each time step.

Results from two-dimensional shock tube tests show that the code captures correctly discontinuities in all three MHD waves families as well as contact discontinuities. The numerical viscosities and resistivity in the code, which are useful in order to understand simulations involving turbulent flows, are estimated through the decay of two-dimensional linear waves. Finally, the robustness of the code in two-dimensions is demonstrated through calculations of the Kelvin-Helmholtz instability and the Orszag-Tang vortex.

*Subject headings:* hydromagnetics - magnetohydrodynamics:MHD - methods:numerical - shock waves

---

## 1. INTRODUCTION

In recent years upwind finite difference schemes based on ideal gas conservation laws have been popular for solving compressible hydrodynamic conservation equations, mainly due to their ability to sharply capture discontinuities and also to the robustness of the schemes. Several examples include Godunov's Scheme (Godunov 1959), MUSCL scheme (Van Leer 1979), Roe Scheme (Roe 1981), TVD scheme (Harten 1983), PPM scheme (Colella & Woodward 1984), and ENO scheme (Harten *et al.* 1987). A comparative review for some of these schemes can be found in Woodward & Colella (1984). Such methods have been very effectively applied to many astrophysical problems (*e.g.*, cosmological hydrodynamics with the TVD scheme (Ryu *et al.* 1993), accretion flow with the ENO scheme (Ishii *et al.* 1993), supernova remnants (Chevalier, Blondin & Emmering 1992) and astrophysical convection (Porter & Woodward 1994) with the PPM scheme).

As important as these hydrodynamical tools have been, there are many applications that fundamentally depend on the presence of magnetic fields and an ionized, conducting character of the medium under investigation. To the extent that these media can be treated through continuum fluid mechanics, this means one must work with magnetohydrodynamic (MHD) equations rather than hydrodynamic equations. But, due to the intrinsic complexity of the MHD flows, the development of numerical techniques to solve MHD equations has been slower than for hydrodynamics. For instance, until now most numerical schemes have been based on methods dependent on artificial viscosity to form shocks(*e.g.*, DeVore 1991; Lind, Payne & Meier 1991; Stone & Norman 1992). Those schemes have been used successfully in astrophysical applications (*e.g.*, Lind *et al.* 1989; Stone & Norman 1994). However, since past experience with fully conservative, high-order upwind hydrodynamic schemes found those to be superior in many applications (Woodward & Colella 1984), it is naturally interesting to extend such schemes to solve MHD conservation equations. Several investigators who have worked in recent years on the development of high-order, conservative, upwind differencing schemes for MHD include Brio & Wu (1988), Zachary & Colella (1992), Zachary, Malagoli, & Colella (1994), and Dai & Woodward (1994a,1994b). Brio & Wu applied Roe's approach to the MHD equations. Zachary and collaborators used the BCT scheme (Bell, Colella, & Tragenstein 1989) to estimate fluxes in MHD conservation equations. Dai & Woodward applied the PPM scheme to MHD. One of the special concerns in developing a scheme for MHD is the fact that the equations may have solutions at which they are no longer strictly hyperbolic, meaning some of the characteristic speeds become locally degenerate.

In a previous work (Ryu & Jones 1995), we described a one-dimensional conservative numerical code to solve the equations for ideal magnetohydrodynamics. It is based on an explicit finite difference scheme on an Eulerian grid, called the Total Variation Diminishing (TVD) scheme (Harten 1983), which is a second-order-accurate extension of the Roe-type upwind scheme. Tests using an extensive set of MHD shock tube problems showed that the one-dimensional code can resolve strong shocks within 2-4 cells; weaker shocks, especially slow shocks are somewhat broader. Contact and tangential discontinuities also require more cells for capture.

This code provides a promising first step to a useful computational tool. However, there are only a very limited number of practical applications that can be studied under the planar symmetry constraint. Most real flows will depend on structures in at least two or, more likely, in three spatial dimensions. Thus, general application of our code requires that it be extended to a multi-dimensional form. In this paper, we describe this extension. In the multi-dimensional code, multiple spatial dimensions are treated through a Strang-type operator splitting (Strang 1968). The constraint of a divergence-free field is enforced exactly by calculating a correction via a gauge transformation in each time step (Brackbill & Barnes 1980).

This scheme should have many interesting application to astrophysical problems. We have already applied the one-dimensional version of our code successfully to the first time dependent study of diffusive cosmic-ray acceleration in oblique MHD shocks (Frank, Jones & Ryu 1994, 1995a). We demonstrated that inclusion of magnetic fields in the fluid dynamics introduces a number of effects, including some that are subtle and not apparent except in a time dependent simulation. With the new multi-dimensional code described here, we can extend that analysis to include such important issues as the evolution of unstable MHD cosmic-ray shocks. Tests we describe below illustrate some of the other kinds of problems that can be addressed with the code. We have underway a study of the nonlinear evolution of the MHD Kelvin-Helmholtz instability (Frank, Jones & Ryu 1995b), for example.

This paper is organized in the following way. In §2, we describe the extension of the one-dimensional code into a multi-dimensional form, including the operator splitting scheme and the correction to preserve $\nabla \cdot \boldsymbol{B} = 0$. In §3, we present the results of numerical tests performed with the two-dimensional version of the code, including shock tube problems, the decay of linear waves, the growth of the Kelvin-Helmholtz instability, and evolution of the Orszag-Tang vortex. Our findings are summarized in §4.

## 2. THE NUMERICAL SCHEME

### 2.1. The MHD Equations

MHD describes the behavior of the combined system of a conducting fluid and magnetic fields in the limit that the displacement current and the separation between ions and electrons are neglected. So, the MHD equations represent coupling of the equations of fluid dynamics with the Maxwell's equations of electrodynamics. Here, we describe a numerical scheme to calculate the evolution of the following ideal MHD equations, where the effects of electrical resistivity, viscosity, and thermal conductivity are dropped,

$$\frac{\partial \rho}{\partial t} + \nabla \cdot (\rho \boldsymbol{v}) = 0, \tag{2.1}$$

$$\frac{\partial \boldsymbol{v}}{\partial t} + \boldsymbol{v} \cdot \nabla \boldsymbol{v} + \frac{1}{\rho}\nabla p - \frac{1}{\rho}\left(\nabla \times \boldsymbol{B}\right) \times \boldsymbol{B} = 0, \tag{2.2}$$

$$\frac{\partial p}{\partial t} + \boldsymbol{v} \cdot \nabla p + \gamma p \nabla \cdot \boldsymbol{v} = 0, \tag{2.3}$$

$$\frac{\partial \boldsymbol{B}}{\partial t} - \nabla \times (\boldsymbol{v} \times \boldsymbol{B}) = 0, \tag{2.4}$$

(see Jackson (1975) for the derivation of the equations). Here, we have chosen units so that factor of $4\pi$ does not appear in the equations. An additional explicit constraint $\nabla \cdot \boldsymbol{B} = 0$ is imposed to account for the absence of magnetic monopoles. Although that constraint is included in the derivation of equation (2.4), it generally cannot be maintained precisely in differenced forms of that equation, even when one works with an exactly conservative scheme as we will outline.

In Cartesian geometry, the above equations are written in conservative form as

$$\frac{\partial \boldsymbol{q}}{\partial t} + \frac{\partial \boldsymbol{F}_x}{\partial x} + \frac{\partial \boldsymbol{F}_y}{\partial y} + \frac{\partial \boldsymbol{F}_z}{\partial z} = 0, \tag{2.5}$$

$$\boldsymbol{q} = \begin{pmatrix} \rho \\ \rho v_x \\ \rho v_y \\ \rho v_z \\ B_x \\ B_y \\ B_z \\ E \end{pmatrix}, \quad \boldsymbol{F}_x = \begin{pmatrix} \rho v_x \\ \rho v_x^2 + p^* - B_x^2 \\ \rho v_x v_y - B_x B_y \\ \rho v_x v_z - B_x B_z \\ 0 \\ B_y v_x - B_x v_y \\ B_z v_x - B_x v_z \\ (E+p^*)v_x - B_x(B_x v_x + B_y v_y + B_z v_z) \end{pmatrix}, \tag{2.6}$$

with $\boldsymbol{F}_y$ and $\boldsymbol{F}_z$ obtained by properly permuting indices. Here, the total pressure and the total energy are given by

$$p^* = p + \frac{1}{2}\left(B_x^2 + B_y^2 + B_z^2\right), \tag{2.7}$$

$$E = \frac{1}{2}\rho\left(v_x^2 + v_y^2 + v_z^2\right) + \frac{p}{\gamma - 1} + \frac{1}{2}\left(B_x^2 + B_y^2 + B_z^2\right). \tag{2.8}$$

With the state vector, $\boldsymbol{q}$, and the flux functions, $\boldsymbol{F}_x(\boldsymbol{q})$, $\boldsymbol{F}_y(\boldsymbol{q})$, and $\boldsymbol{F}_z(\boldsymbol{q})$, the Jacobian matrices, $\boldsymbol{A}_x(\boldsymbol{q}) = \partial \boldsymbol{F}_x/\partial \boldsymbol{q}$, $\boldsymbol{A}_y(\boldsymbol{q}) = \partial \boldsymbol{F}_y/\partial \boldsymbol{q}$, and $\boldsymbol{A}_z(\boldsymbol{q}) = \partial \boldsymbol{F}_z/\partial \boldsymbol{q}$, are formed. The system of equations is called *hyperbolic* if all the eigenvalues of the Jacobian matrices are real and distinct and the corresponding set of right eigenvectors is complete (Jeffrey & Taniuti 1964). The MHD equations form a *non-strictly hyperbolic* system, meaning that some eigenvalues may coincide at some points.

### 2.2. The One-Dimensional Code

The procedure to build a one-dimensional MHD code based on the TVD scheme was described in detail in our previous paper (Ryu & Jones 1995). Here, we briefly summarize it and mention optimizations for the multi-dimensional extension. To start, we consider a plane-symmetric, one-dimensional flow exhibiting variation along the $x$-direction. The first step to build the code is to find the eigenvalues and the right and left eigenvectors of the Jacobian matrix, $\boldsymbol{A}_x(\boldsymbol{q})$. The seven eigenvalues $a_1, \cdots, a_7$ in non-increasing order are

$$a_{1,7} = v_x \pm c_f, \qquad a_{2,6} = v_x \pm c_a, \qquad a_{3,5} = v_x \pm c_s, \qquad a_4 = v_x, \tag{2.9}$$

where $c_f$, $c_a$, $c_s$ are the fast, Alfvén, and slow characteristic speeds. The quantities $a_1, \cdots, a_7$ represent the seven speeds with which information is propagated locally by three MHD wave families and an entropy mode. The three characteristic speeds are expressed as

$$c_{f,s} = \left[ \frac{1}{2} \left\{ a^2 + \frac{B_x^2 + B_y^2 + B_z^2}{\rho} \pm \sqrt{\left( a^2 + \frac{B_x^2 + B_y^2 + B_z^2}{\rho} \right)^2 - 4a^2 \frac{B_x^2}{\rho}} \right\} \right]^{\frac{1}{2}}, \tag{2.10}$$

and

$$c_a = \sqrt{\frac{B_x^2}{\rho}}, \tag{2.11}$$

with the sound speed,

$$a = \sqrt{\gamma \frac{p}{\rho}}. \tag{2.12}$$

The corresponding eigenvectors are given, for example, in Jeffrey & Taniuti (1964). However, near a point where either $B_x = 0$ or $B_y = B_z = 0$, the eigenvectors are not well defined, with elements becoming singular. By renormalizing the eigenvectors, the singularities can be removed. The renormalized eigenvectors are given in Brio & Wu (1988) and Ryu & Jones (1995).

In a code based on the TVD scheme, the physical quantities are referred to the cell centers while the fluxes are computed on the cell boundaries. Implementation of Roe's linearization technique would result in a particular averaged form of the physical quantities on the cell boundaries (Roe 1981). However, as pointed out by Brio & Wu (1988), it is not possible to derive this particular analytic form of the averaged quantities in MHD for general cases with an adiabatic index $\gamma \neq 2$. Instead, we modify Roe's scheme and use $\rho_{i+\frac{1}{2}}$, $v_{x,i+\frac{1}{2}}$, $v_{y,i+\frac{1}{2}}$, $v_{z,i+\frac{1}{2}}$, $B_{y,i+\frac{1}{2}}$, $B_{z,i+\frac{1}{2}}$, $p^*_{i+\frac{1}{2}}$ on the cell boundaries with the arithmetic averages at $i$ and $i + 1$. Then, other quantities like momentum, gas pressure, total energy, etc are calculated by combining those quantities. Our tests for cases with $\gamma = 2$ indicated that the above simple averaging would do just as well when compared to the full implementation of Roe's linearization technique, as already suggested for the original hydrodynamic code by Harten (1983).

The state vector $\boldsymbol{q}^n$ at the time step $n$ is updated by calculating the modified flux $\bar{\boldsymbol{f}}_x$ at the cell boundaries as follows:

$$L_x \boldsymbol{q}_i^n = \boldsymbol{q}_i^n - \frac{\Delta t^n}{\Delta x} (\bar{\boldsymbol{f}}_{x,i+\frac{1}{2}} - \bar{\boldsymbol{f}}_{x,i-\frac{1}{2}}), \tag{2.13}$$

$$\bar{f}_{x,i+\frac{1}{2}} = \frac{1}{2}\left[F_x(q_i^n) + F_x(q_{i+1}^n)\right] - \frac{\Delta x}{2\Delta t^n}\sum_{k=1}^{7}\beta_{k,i+\frac{1}{2}}R_{k,i+\frac{1}{2}}^n, \qquad (2.14)$$

$$\beta_{k,i+\frac{1}{2}} = Q_k\left(\frac{\Delta t^n}{\Delta x}a_{k,i+\frac{1}{2}}^n + \gamma_{k,i+\frac{1}{2}}\right)\alpha_{k,i+\frac{1}{2}} - (g_{k,i} + g_{k,i+1}), \qquad (2.15)$$

$$\alpha_{k,i+\frac{1}{2}} = L_{k,i+\frac{1}{2}}^n \cdot (q_{i+1}^n - q_i^n), \qquad (2.16)$$

$$\gamma_{k,i+\frac{1}{2}} = \begin{cases} \frac{g_{k,i+1} - g_{k,i}}{\alpha_{k,i+\frac{1}{2}}}, & \text{for } \alpha_{k,i+\frac{1}{2}} \neq 0 \\ 0, & \text{for } \alpha_{k,i+\frac{1}{2}} = 0 \end{cases}, \qquad (2.17)$$

$$g_{k,i} = \text{sign}(\tilde{g}_{k,i+\frac{1}{2}})\max\left[0, \min\left\{|\tilde{g}_{k,i+\frac{1}{2}}|, \tilde{g}_{k,i-\frac{1}{2}}\text{sign}(\tilde{g}_{k,i+\frac{1}{2}})\right\}\right], \qquad (2.18)$$

$$\tilde{g}_{k,i+\frac{1}{2}} = \frac{1}{2}\left[Q_k(\frac{\Delta t^n}{\Delta x}a_{k,i+\frac{1}{2}}^n) - (\frac{\Delta t^n}{\Delta x}a_{k,i+\frac{1}{2}}^n)^2\right]\alpha_{k,i+\frac{1}{2}}, \qquad (2.19)$$

$$Q_k(\chi) = \begin{cases} \frac{\chi^2}{4\varepsilon} + \varepsilon, & \text{for } |\chi| < 2\varepsilon \\ |\chi|, & \text{for } |\chi| \geq 2\varepsilon \end{cases}, \qquad (2.20)$$

$$\varepsilon = \begin{cases} 0.3, & \text{for } k = 1 \text{ and } 7 \\ 0.1, & \text{for } k = 2 \text{ and } 6 \\ 0.45, & \text{for } k = 3 \text{ and } 5 \\ 0.0 & \text{for } k = 4 \end{cases}. \qquad (2.21)$$

Here, $a_k$, $R_k$, $L_k$ are the seven eigenvalues in (2.9) and the corresponding right and left eigenvectors. The time step $\Delta t^n$ is restricted by the usual Courant condition for the stability, $\Delta t^n = C_{\text{cour}}\Delta x/\text{Max}(|v_{x,i+\frac{1}{2}}^n| + c_{f,i+\frac{1}{2}}^n)$ with $C_{\text{cour}} < 1$. Typically we use $C_{\text{cour}} = 0.6$, although values up to 0.9 seems to be sufficient for most calculations. Note that the $\varepsilon$ and $C_{\text{cour}}$ values in (2.20) as optimized for the multi-dimensional code are different from those recommended for the one-dimensional code (Ryu & Jones 1995). Our tests showed that procedures to steepen contact discontinuities and rotational discontinuities similar to those suggested in the original TVD paper by Harten (1983) do not work very well in this context. They produce spurious numerical oscillations that can significantly degrade the solution. So, we do not include *contact steepener* and *rotational steepener* routines in our multi-dimensional MHD code.

### 2.3. The Multi-Dimensional Extension

The multi-dimensional extension has been done through a Strang-type directional splitting (Strang 1968). In each time step, multi-dimensional derivatives are *split* into a set of one-dimensional derivatives, with variations in other directions ignored temporarily. Then, each row and column in the grid is treated as if it were a one-dimensional problem. Updating the flow quantities along each row is done using the one-dimensional code described in the previous section. The parallel (to the direction of the row) component of magnetic field is kept constant and only the perpendicular component is updated. One complete time step updating the full state vector $\boldsymbol{q}^n$ to $\boldsymbol{q}^{n+1}$ in each cell is composed of updating it along two or three directions, as appropriate. For instance, in three-dimensional Cartesian geometry, the state vector is updated along $x$, $y$, and $z$-directions, so

$$\boldsymbol{q}^{n+1} = L_z L_y L_x \boldsymbol{q}^n. \tag{2.22}$$

In order to maintain second-order accuracy, the order of directional passes is permuted by the Strang-type prescription: $L_z L_y L_x$, $L_x L_y L_z$, $L_x L_z L_y$, $L_y L_z L_x$, $L_y L_x L_z$, and then $L_z L_x L_y$, for example. The time step is restricted to satisfy the Courant condition along each row in three directions. It is calculated at the start of the above permuting sequence and used through one complete sequence.

### 2.4. Ensuring $\nabla \cdot \boldsymbol{B} = 0$

The condition $\nabla \cdot \boldsymbol{B} = 0$ is, of course, a necessary initial constraint in multi-dimensional MHD flows and should be preserved during their evolution. While the differential MHD equations formally ensure $\nabla \cdot \boldsymbol{B} = 0$, the numerical errors due to discretization and operator splitting can lead to non-zero divergence over time. Physically, this is due to the fact that even though schemes such as this one are exactly conservative of the fluxes in equation (2.6) on the cell boundaries crossing each directional pass, nothing maintains a constant magnetic flux in the *Gauss's Law* sense across the entire surface of each cell. That is, nothing forces conservation of magnetic charge over a finite cell during the entire time step. This error usually grows exponentially during the computations, causing an artificial force parallel to the magnetic field and destroying the correct dynamics of the flows (see *e.g.*, Schmidt-Voigt 1987).

Zachary, Malagoli, & Colella (1994) described that, by using the modified form of the MHD equations originally suggested in Brackbill & Barnes (1980), the error in $\nabla \cdot \boldsymbol{B}$ could be kept small enough in their code that their tests showed no discernible differences with and without the correction for non-zero $\nabla \cdot \boldsymbol{B}$. So they abandoned the correction in their code. However, since the modified form is not suitable for our code and it does not remove $\nabla \cdot \boldsymbol{B}$ completely anyway, we enforce $\nabla \cdot \boldsymbol{B} = 0$ explicitly by a simple transformation. First we solve for the potential, $\phi$, defined by the Poisson equation

$$\nabla^2 \phi + \nabla \cdot \boldsymbol{B} = 0. \tag{2.23}$$

Here, $\boldsymbol{B}$ is the updated magnetic field obtained by the procedure already outlined. Then we compute the corrected magnetic field, defined as $\boldsymbol{B}^c = \boldsymbol{B} + \nabla \phi$, for which $\nabla \cdot \boldsymbol{B}^c = 0$, as required.

The difference form of the above Poisson equation in two-dimensional Cartesian geometry is

$$\frac{\phi_{i+2,j} - 2\phi_{i,j} + \phi_{i-2,j}}{4(\Delta x)^2} + \frac{\phi_{i,j+2} - 2\phi_{i,j} + \phi_{i,j-2}}{4(\Delta y)^2}$$
$$= -\left(\frac{B_{x,i+1,j} - B_{x,i-1,j}}{2\Delta x} + \frac{B_{y,i,j+1} - B_{y,i,j-1}}{2\Delta y}\right), \quad (2.24)$$

To solve the above difference equation, we employ a suitably selected rapid technique. On a Cartesian grid, for example, we can take advantage of the fact from the Fourier transform *derivative theorem* that FFTs (fast Fourier transforms) can provide an exact solution to $\phi_{i,j}$ from equation (2.23) if the field structure is periodic on some space (*e.g.*, Bracewell 1986). Even if the magnetic field is not periodic on the computational box of interest, it is often possible to create such a structure for this transformation by doubling the box size *for this step only* and choosing an appropriate symmetry within the extended space. According to our tests, the cost to enforce $\nabla \cdot \boldsymbol{B} = 0$ explicitly is reasonable, taking $\lesssim 5\%$ of the total CPU time in cases with periodic boundaries on the computational space. With others, like reflecting or continuous boundaries, that require an extended space the cost to enforce $\nabla \cdot \boldsymbol{B} = 0$ is $\lesssim 30\%$ of the total CPU time.

The present version of the multi-dimensional MHD TVD code runs at $\sim 400$ MFlops on a single Cray C90 processor, although it is not specifically optimized for the machine. This corresponds to an update rate of $\sim 2 \times 10^5$ cells/sec for the one-dimensional version and $\sim 10^5$ cells/sec for the two-dimensional version.

## 3. TESTS

In this section we briefly describe several tests we have carried out to determine the characteristics of the code and its suitability for practical application. We aim to examine the ability of the code to handle all three MHD wave families as well as its performance in computing complex flows. In addition we will estimate the effects of numerical viscosity and electrical resistivity. For each of these tests we assume an adiabatic index, $\gamma = 5/3$ and a Courant constant, $C_{cour} = 0.6$.

### 3.1. Shock Tube tests

Extensive tests of the code have been done with MHD shock tube problems, using a two-dimensional Cartesian grid and various orientations of the structures with respect to the grid. Here we describe tests performed in a two-dimensional box with $x = [0, 1]$ and $y = [0, 1]$, where structures like discontinuities and waves propagate along the diagonal line joining $x = y = 0$ and $x = y = 1$. We present two tests. One includes only two ($x$ and $y$) components of magnetic field and velocity, so that they are confined in the computational plane. The second includes all three components. The results from the numerical calculations are marked with dots in Figures 1 and 2. They can be compared with analytic solutions from the nonlinear Riemann solver described in Ryu and Jones (1995), plotted here with lines. Structures are measured along the diagonal line joining $x = y = 0$ and $x = y = 1$. The plotted quantities are density, gas pressure, total energy, $v_\parallel$ (velocity parallel to the diagonal line; i.e., parallel to the wave normal), $v_\perp$ (velocity perpendicular to the diagonal line but still in the computational plane), $v_z$ (velocity in the direction out of plane), and the analogous magnetic field components, $B_\parallel$, $B_\perp$, and $B_z$.

The first test in Fig. 1 has been done with two-dimensional magnetic field and velocity vectors in the $x-y$ plane. In this case the sign of the perpendicular magnetic field ($B_\perp$) remains unchanged across the structures. The initial left state is $(\rho, v_\parallel, v_\perp, v_z, B_\perp, B_z, E) = (1, 10, 0, 0, 5/\sqrt{4\pi}, 0, 20)$ and the initial right state is $(1, -10, 0, 0, 5/\sqrt{4\pi}, 0, 1)$, with $B_\parallel = 5/\sqrt{4\pi}$. Fig. 1a in Ryu & Jones (1995) illustrates a solution to this problem with the one-dimensional MHD TVD code. The calculation has been done using $256 \times 256$ cells, and plots correspond to time $t = 0.08\sqrt{2}$. The structure is bounded by a left and right facing fast shock pair. There are also a left facing slow rarefaction, a right facing slow shock and a contact discontinuity. Three anomalous points visible in the $v_\perp$ plot are due to the error induced in the subtraction of two big numbers to get a small number. Note that $v_\perp$ has been calculated with $v_\perp = (-v_x + v_y)/\sqrt{2}$ and $v_x$ and $v_y$ are, at least, an order of magnitude larger than $v_\perp$ in preshock regions. As expected from the two-dimensional nature of the magnetic field and velocity structure, there is no rotational discontinuity. In this test the fast shocks are strong, with a large parallel velocity jump, $[v_\parallel]$. The density, parallel velocity and pressure jumps in the slow shock are weak, so that this feature is mainly apparent through the jumps in the tangential velocity, $[v_\perp]$ and tangential magnetic field, $[B_\perp]$. The capture of shocks and the contact discontinuity here are virtually the same as with the one-dimensional code.

The second test in Fig. 2 involves all three components of both magnetic field and velocity, with the magnetic field plane rotated across the initial discontinuity (such calculations are often described as 2 + 1/2 dimensional). This test is also presented as Fig. 2a in Ryu & Jones (1995). The initial left state is $(\rho, v_\parallel, v_\perp, v_z, B_\perp, B_z, E) = (1.08, 1.2, 0.01, 0.5, 3.6/\sqrt{4\pi}, 2/\sqrt{4\pi}, 0.95)$ and the initial right state is $(1, 0, 0, 0, 4/\sqrt{4\pi}, 2/\sqrt{4\pi}, 1)$, with $B_\parallel = 2/\sqrt{4\pi}$. Again the calculation has been done using $256\times 256$ cells, and plots correspond to time $t = 0.2\sqrt{2}$. Fast shocks, rotational discontinuities, and slow shocks propagate from each side of the contact discontinuity. Here, the rotation of the magnetic field across the initial discontinuity generates two rotational discontinuities. Again, the structures are captured by the two-dimensional code very similarly to what we found with the one-dimensional code.

We also carried out many of the other shock tube tests described in Ryu & Jones with this code, including rarefactions of both fast and slow wave families. Generally the solutions are comparable to those seen there. In summary, then, results from multi-dimensional shock tube tests show that the code captures correctly discontinuities and rarefaction waves in all three MHD wave families, as well as that carried with the entropy mode.

## 3.2. Decay of Linear Waves

Our conservation equations (2.5-2.6) refer to ideal MHD without the effects of viscosity, electrical resistivity, and thermal conductivity. However, in numerical calculations on a discrete grid, diffusion of magnetic field and momentum across cell boundaries is unavoidable and introduces effective *numerical* electrical resistivity and viscosity. Energy diffusion also occurs and produces *numerical* thermal conductivity. But, the effects of the numerical thermal conductivity are mostly small compared to those of the numerical viscosity and electrical resistivity.

In order to estimate the properties of the numerical resistivity and viscosity in the multi-dimensional MHD code, we have followed the decay of two-dimensional Alfvén, fast, and slow waves in numerical calculations and compared their decay rates to the predicted rates in a viscous, resistive fluid. The MHD equations for a viscous, resistive fluid can be written as

$$\frac{\partial \rho}{\partial t} + \nabla \cdot (\rho \boldsymbol{v}) = 0, \tag{3.1}$$

$$\frac{\partial \boldsymbol{v}}{\partial t} + \boldsymbol{v} \cdot \nabla \boldsymbol{v} + \frac{1}{\rho}\nabla p - \frac{1}{\rho}(\nabla \times \boldsymbol{B}) \times \boldsymbol{B} = \frac{1}{\rho}\partial_k \sigma_{ik}, \tag{3.2}$$

$$\frac{\partial p}{\partial t} + \boldsymbol{v} \cdot \nabla p + \gamma p \nabla \cdot \boldsymbol{v} = (\gamma - 1)\sigma_{ik}\partial_k v_i, \tag{3.3}$$

$$\frac{\partial \boldsymbol{B}}{\partial t} - \nabla \times (\boldsymbol{v} \times \boldsymbol{B}) = \eta \nabla^2 \boldsymbol{B}. \tag{3.4}$$

In these equations,

$$\sigma_{ik} = \mu\left(\partial_k v_i + \partial_i v_k - \frac{2}{3}\delta_{ik}\nabla \cdot \boldsymbol{v}\right) + \zeta \delta_{ik} \nabla \cdot \boldsymbol{v} \tag{3.5}$$

is the viscosity tensor, where $\mu$ and $\zeta$ are the dynamic shear and bulk viscosity, respectively, $\eta$ is the electrical resistivity, and $\partial_k = \partial/\partial x_k$.

In the test of the decay of Alfvén (shear) waves, we have used a standing wave formed along the grid diagonal with initial conditions

$$\delta v_z = v_{\text{amp}} c_a \sin\left(k_x x + k_y y\right),$$

$$\delta \rho = \delta p = \delta v_x = \delta v_y = \delta B_x = \delta B_y = \delta B_z = 0, \tag{3.6}$$

in a stationary background with $\rho_o = 1$, $p_o = 1$, and $\boldsymbol{B} = B_o \hat{\boldsymbol{x}}$ with $B_o = 1$. This gives characteristic speeds, $a = 1.291$ and $c_a = 0.7071$. The calculations have been done in a square periodic box with size $L = 1$ using $8 \times 8$, $16 \times 16$, $32 \times 32$, $64 \times 64$, $128 \times 128$, and $256 \times 256$ cells. The $x$ and $y$-wavenumbers have been set to $k_x = k_y = 2\pi/L$, so the total wavenumber $k = \sqrt{k_x^2 + k_y^2} = \sqrt{2}(2\pi/L)$, and the initial peak amplitude has been set to $v_{\text{amp}} = 0.1$. The predicted complex frequency of the wave is

$$\omega = \frac{i}{2}\left(\frac{\mu}{\rho_o} + \eta\right)k^2 \pm c_a k \sqrt{1 - \frac{1}{4c_a^2}\left(\frac{\mu}{\rho_o} - \eta\right)^2 k^2}, \tag{3.7}$$

so the decay rate is

$$\Gamma_a = \frac{1}{2}\left(\frac{\mu}{\rho_o} + \eta\right) k^2. \tag{3.8}$$

The effect of the shear viscosity and the electrical resistivity on Alfvénic turbulence of scale, $L$, is measured through a *Reynolds number* defined by

$$R_a \equiv c_a L \left/ \left[\frac{1}{2}\left(\frac{\mu}{\rho_o} + \eta\right)\right]\right. = \frac{8\pi^2 c_a}{L\Gamma_a}. \tag{3.9}$$

When the Reynolds number corresponding to $L$ is large the fluid should be able to maintain on that scale a turbulent flow whose properties are not controlled by the properties of the viscosity and resistivity.

In a similar fashion we have set up tests of the decay of compressive waves. Each involves standing waves formed on the grid diagonal. For the fast wave we used initial conditions

$$\delta v_x = v_{\text{amp}} c_f \sin(k_x x + k_y y), \qquad \delta v_y = \frac{c_f^2}{c_f^2 - B_o^2/\rho_o} v_{\text{amp}} c_f \sin(k_x x + k_y y),$$

$$\delta\rho = \delta p = \delta v_z = \delta B_x = \delta B_y = \delta B_z = 0. \tag{3.10}$$

For the slow wave we set up initial conditions

$$\delta v_x = v_{\text{amp}} c_s \sin(k_x x + k_y y), \qquad \delta v_y = \frac{c_s^2}{c_s^2 - B_o^2/\rho_o} v_{\text{amp}} c_s \sin(k_x x + k_y y),$$

$$\delta\rho = \delta p = \delta v_z = \delta B_x = \delta B_y = \delta B_z = 0. \tag{3.11}$$

The same background configuration and computational box have been used as in the case of Alfvén waves. For comparison, the characteristic wave speeds are $c_f = 1.518$ and $c_s = 0.6013$. In compressive waves with a large wave amplitude, nonlinear effects become important (e.g., steepening), so a smaller initial peak amplitude, $v_{\text{amp}} = 10^{-4}$, has been used. The derivation of the predicted frequency and decay rate of the fast and slow waves in a viscous, resistive fluid is complicated. The analytic expressions can be obtained only up to the first order of $\mu$, $\zeta$, and $\eta$, in the limit

$$\frac{\mu}{\rho_o}k, \quad \frac{\zeta}{\rho_o}k, \quad \text{and} \quad \frac{\eta}{\rho_o}k \ll c_f, \quad c_a, \quad c_s, \quad \text{or} \quad a, \tag{3.12}$$

a condition usually satisfied. The resulting decay rate for fast waves is

$$\Gamma_f = \frac{k^2}{2\left(c_f^2 - c_s^2\right)} \left[c_f^2\left(\frac{7}{3}\frac{\mu}{\rho_o} + \frac{\zeta}{\rho_o} + \eta\right) - \frac{B_o^2}{\rho_o}\frac{\mu}{\rho_o} - c_a^2\left(\frac{1}{3}\frac{\mu}{\rho_o} + \frac{\zeta}{\rho_o}\right) - a^2\left(\frac{\mu}{\rho_o} + \eta\right)\right], \tag{3.13}$$

while the decay rate for slow waves is

$$\Gamma_s = \frac{k^2}{2\left(c_f^2 - c_s^2\right)} \left[\frac{B_o^2}{\rho_o}\frac{\mu}{\rho_o} + c_a^2\left(\frac{1}{3}\frac{\mu}{\rho_o} + \frac{\zeta}{\rho_o}\right) + a^2\left(\frac{\mu}{\rho_o} + \eta\right) - c_s^2\left(\frac{7}{3}\frac{\mu}{\rho_o} + \frac{\zeta}{\rho_o} + \eta\right)\right]. \tag{3.14}$$

As in the case of Alfvén waves, we define *Reynolds numbers*

$$R_f \equiv \frac{8\pi^2 c_f}{L\Gamma_f} \tag{3.15}$$

for the decay of fast waves

$$R_s \equiv \frac{8\pi^2 c_s}{L\Gamma_s} \tag{3.16}$$

and for the decay of slow waves.

Fig. 3 shows an example from a $32 \times 32$ cell test of the decay of a linear wave, an Alfvén wave in this case. In most calculations, including this one, the decay pattern fits well to an exponential form, giving well-defined decay rates.

In Fig. 4, the normalized decay rates and Reynolds numbers in the tests using Alfvén, fast, and slow waves are plotted as a function of the number of cells spanning the length $L$. Good fits of the form $\Gamma \propto n_{\text{cells}}^{-2}$ and $R \propto n_{\text{cells}}^2$ are possible, confirming that our code has second-order accuracy. Ryu & Goodman (1994) carried out analogous tests for the hydrodynamic TVD code by examining decay of two-dimensional sound waves and plane shear flows (Ryu & Goodman 1994). Those tests also produce a numerical Reynolds number that scales as $R \propto n_{\text{cells}}^2$. In the case with a relatively strong field examined here (the plasma $\beta \equiv p_o/(B_o^2/2) = 2$), the numerical Reynolds number that applies to a particular number of cells is, on average, roughly 20 -30 times smaller than that in the hydrodynamic TVD code.

However, the amount of numerical dissipation also depends on background configuration, especially on background field strength in MHD codes. So we have followed the decay of a fast wave in a background with different magnetic field strength ($B_o$). The fast wave becomes a sound wave, if $B_o = 0$. We have used $B_o = 10$, 3, 1, 0.3, 0.1, 0.03, 0.01 0.003, 0.001, and 0 (so $\beta = 2 \times 10^{-2}$, $\cdots$, $2 \times 10^6$, and $\infty$) with $128 \times 128$ cells. For $B_o = 0$, the hydrodynamic TVD code has been used (Harten 1983; Ryu and Goodman 1994). Except $B_o$, the background configuration is same as that used in the test in Fig. 4. Fig. 5 shows the resulting normalized decay rate and Reynolds number as functions of $\beta$. Here, the normalized decay rate is biggest around $\beta = 1$, which is the case of the test in Fig. 4. The reason why the normalized decay rate is decreased with decreasing $\beta$ for $\beta \lesssim 1$ is because the fast wave speed increases faster than the decrease in the decay rate. The decay rate in the MHD TVD code converges as $\beta$ goes $\infty$. But still it is significantly larger than that in the hydrodynamic TVD code. This is because the MHD TVD code does not reduce exactly to the hydrodynamic TVD code even $B = 0$. There are 7 characteristic fields in MHD, and 5 characteristic fields for hydrodynamics are recovered by combining them properly, as was described in Ryu and Jones (1995). This procedure puts in some additional numerical diffusion.

The above tests mean that to reach an inertial range simulations of MHD turbulence, especially with strong field ($\beta \sim 1$), would need to resolve fine structures with more cells than those of hydrodynamic turbulence. That could be an important aspect of such calculations. We suspect that this will be a common property of many numerical schemes, but are not aware of any previously published tests that address the issue. Computations primarily concerned with MHD shocks, however, can be conducted with similar resolutions as those for hydrodynamic shocks. Numerical dissipation measured in this subsection influences smooth flows only, and shocks, especially strong shocks, are resolved in the MHD TVD code as sharply as in the hydrodynamic TVD code.

### 3.3. The Kelvin-Helmholtz Instability

To evaluate its potential for practical problems, it is essential to test a numerical method's ability to simulate complex, fully two-dimensional flows. We present two sample test problems that include some challenging properties, such as strong shear and obliquely intersecting shocks. The first test involves an unstable, compressible shear layer; *i.e.*, the compressible MHD Kelvin-Helmholtz instability. To make this test as *clean* and decisive as possible, we have chosen as initial conditions a smoothly sheared flow with linear, eigenmode perturbations that have precisely known solutions (Miura & Pritchett 1982).

Initial equilibrium configurations exist in which the magnetic field lines lie in planes of constant velocity. For perturbations of this with associated wave vectors parallel to the equilibrium velocity field, the dynamical role of the magnetic field depends on the angle between the field and the wave vector. If the field is perpendicular to the wave vector (that is, out of the computational plane in a two-dimensional simulation) the field can only be compressed in perturbations, not *stretched*, so the perturbation behaves qualitatively like a gasdynamic flow. A much more interesting case for our purposes begins with the magnetic field parallel to the flow, since perturbations then bend and stretch field lines. We have actually tested both cases, but present here only the latter case.

Following Miura & Pritchett (1982), we can describe a perturbed equilibrium state as:

$$\rho(x,y,t) = \rho_o + \delta\rho(x,y,t),$$
$$v_x(x,y,t) = V_{xo}(y) + \delta v_x(x,y,t),$$
$$v_y(x,y,t) = \delta v_y(x,y,t),$$
$$B_x(x,y,t) = B_{xo} + \delta B_x(x,y,t), \quad (3.17)$$
$$B_y(x,y,t) = \delta B_y(x,y,t),$$
$$p(x,y,t) = p_o + \delta p(x,y,t),$$
$$v_z(x,y,t) = B_z(x,y,t) = 0,$$

where the equilibrium velocity profile $V_{xo}(y)$ is given by

$$V_{xo}(y) = -\frac{V_o}{2}\tanh(\frac{y - 0.5L}{h}), \quad (3.18)$$

and the equilibrium gas density and pressure along with the magnetic field are uniform. The length, $h$, measures the thickness of the shear layer.

The flow is defined in a square computational box with $x = [0, L]$, $y = [0, L]$. The box is assumed to be periodic in the $x$-direction, and has reflecting $y$-boundaries. Each of the perturbed quantities takes a form

$$\delta f(x, y, t) = f(y) \exp(i k_x x + i \omega t), \tag{3.19}$$

where $k_x = 2\pi/L$ and $f(y)$ is a complex function found by numerically integrating the linearized versions of equations (2.1-2.4) (see Miura & Pritchett (1982)). In practice it is most convenient to define these quantities in terms of the perturbed total pressure, $\delta p^*$. To obtain that quantity we integrated from one of the $y$-boundaries to the midplane and assumed antisymmetric and symmetric continuations through the midplane for the real and imaginary parts, respectively. Further details of the method, including boundary conditions can be obtained from the Miura & Pritchett paper. From symmetry the real part of the frequency $\omega$ is zero in the computational reference frame. The imaginary part, or growth rate, $\Gamma$, can be obtained by iteration on the solution. For our purposes it was sufficient to use values for $\Gamma$ obtained in this fashion and published in graphical form by Miura & Pritchett.

Our test assumes $\rho_O = 1.0$; $p_O = 0.6$ (sound speed, $a = 1$); $B_x = 0.4$ (Alfvén speed, $c_a = 0.4$); $V_O = 1.0$. Thus, the sonic Mach number of the flow is $M_S = 1$ and the Alfvénic Mach number is $M_A = 2.5$. We chose $h = L/(8\pi)$, which places the $y$ boundaries far enough from the strongly sheared flow to produce minimal boundary effects during the linear phase of the instability. The linear growth rate for this mode is $\Gamma = 0.68 V_O/L$. For the test shown we began with a perturbation such that $|\delta p^*(y = L)| = 0.001 p_O^*$. That carries peak perturbations $|\delta p^*| = 0.058 p_O^*$ and $|\delta v_y| = 0.07 V_O$.

Figure 5 illustrates the early growth of the perturbation as measured by the root mean squared transverse velocity, $\langle v_y^2 \rangle^{1/2}$, normalized to the initial value. Results determined from simulations with three resolutions, $128 \times 128$, $256 \times 256$, and $512 \times 512$ cells are shown along with the predicted, exponential growth curve. After an initial transient, the simulated instability grows at the expected rate for $t \sim 1/\Gamma$, before saturation becomes significant. The transverse velocity saturates at larger values in the higher resolution calculations presumably because the effects of numerical dissipation are reduced.

In Figure 6 we present images of the $512 \times 512$ simulation at $t = 1.4/\Gamma$. Dynamically this time is close to the earlier time shown in Fig. 4 of Miura (1984) for a simulation of the same problem. Flow properties are apparently very similar although ours is of higher resolution and shows much finer detail. The quantities shown in Fig. 6, here, are (from left to right, top to bottom): $\rho$; $p_m$ (magnetic pressure); $p$ (gas pressure); magnetic field lines (contours of vector potential $A_z$). The simulation figure has captured the development of an intense magnetic field region left of center on the grid and shows clearly how gas is excluded from that structure. Not obvious in this view, but apparent in displays of the flow velocity is a nascent vortex just to the right of center.

### 3.4. The Orszag-Tang Vortex

In the final test, we have followed the formation of the compressible Orszag-Tang vortex. The problem was originally studied by Orszag & Tang (1979) in the context of incompressible MHD turbulence, and was later extended by Dahlburg & Picone (1989) and Picone & Dahlburg (1991) to compressible MHD turbulence. Zachary, Malagoli, & Colella (1994) applied this problem to test their code. We have used this problem to demonstrate the robustness of our code in handling problems involving multiple shocks and their collisions and also to compare our code, at least qualitatively, with the code by Zachary, Malagoli, & Colella (1994) in a complex, albeit two-dimensional, MHD flow.

The test has been set up with the initial velocity and magnetic field given by

$$\boldsymbol{v} = v_O \left[ -\sin(2\pi y)\hat{\boldsymbol{x}} + \sin(2\pi x)\hat{\boldsymbol{y}} \right], \tag{3.20}$$

$$\boldsymbol{B} = B_O \left[ -\sin(2\pi y)\hat{\boldsymbol{x}} + \sin(4\pi x)\hat{\boldsymbol{y}} \right], \tag{3.21}$$

with $v_O = 1$ and $B_O = 1/\sqrt{4\pi}$. Both the velocity and magnetic field contain similar x-points, but they have different modal structures. The background density and pressure have been assumed to be uniform with values fixed by

$$M^2 \equiv \frac{v_O}{(\gamma p_O/\rho_O)} = 1, \tag{3.22}$$

$$\beta \equiv \frac{p_O}{(B_O^2/2)} = \frac{10}{3}, \tag{3.23}$$

with $\gamma = 5/3$. The calculation has been done in a periodic box with $x = [0,1]$ and $y = [0,1]$ using $256 \times 256$ cells.

Fig. 7 shows the gray scale images of gas pressure, magnetic pressure, as well as compression, $\nabla \cdot \boldsymbol{v}$, and vorticity, $(\nabla \times \boldsymbol{v})_z$ at time $t = 0.48$. The overall flow structure indicates that the gas pressure and the magnetic pressure are anti-correlated resulting in the smoother total pressure, except in the postshock flows where both the gas and magnetic pressures increase sharply. The $\nabla \cdot \boldsymbol{v}$ plot traces the loci of many shocks, which are interacting with each other. Usually shocks, especially strong shocks, are resolved sharply within $2-3$ cells. The $(\nabla \times \boldsymbol{v})_z$ plot demonstrates the existence of many fine sheared structures that are not visible in other plots. Even though we have neither used exactly the same initial conditions nor made plots at exactly the same epoch, the images show that the overall shape and dynamics match closely with those in Dahlburg & Picone (1989) and Zachary, Malagoli, & Colella (1994).

## 4. Summary

In this paper, we have described a multi-dimensional numerical code to solve the ideal MHD equations. It is based on an explicit and conservative finite difference scheme on an Eulerian grid, called the TVD scheme (Harten 1983). The TVD scheme is a second-order-accurate extension of the Roe's upwind scheme (Roe 1981). We previously developed and tested a one-dimensional version of the code (Ryu & Jones 1995) and carried out some successful applications, as described in §1. The multi-dimensional extension has been done through a Strang-type directional splitting (Strang 1968). The constraint $\nabla \cdot \boldsymbol{B} = 0$ has been enforced exactly by calculating a correction via a gauge transformation at each time step.

We have carried out a number of tests of the method using a two-dimensional Cartesian grid to demonstrate its robustness. From shock tube tests we find that the method successfully captures all types of MHD discontinuities. It captures strong fast mode shocks generally within 2-3 numerical cells, while weak, slow mode shocks may spread over more cells. Other discontinuities, such as rotational discontinuity, contact discontinuity, and tangential discontinuity, generally require more cells for containment too. We tested the diffusive and dissipative properties of the code by examining the decay of three types of linear MHD waves. From this we conclude that to bring turbulent flows within an *inertial*, non-dissipative range requires that important structures exceed $\sim 50$ cells. To examine the ability of the code to correctly follow complex multi-dimensional MHD flows we carried out simulations of the evolution of an unstable shear flow and also followed the evolution of the Orszag-Tang vortex. In both cases our results agree with previously published behaviors.

The method is reasonably fast. Our current version of the code runs at about 400 MFlops on a Cray C90 processor, so that about $10^5$ two-dimensional zones can be updated per second.

These properties seem to make the code an attractive method for solving practical problems involving MHD flows in astrophysical contexts. It is reasonably accurate and robust with small enough numerical diffusivity that meaningful solutions to a variety of two-dimensional problems ought to be achievable with it using only modest computational resources by today's standards.

The work by DR was supported in part by the Non-Directed Research Fund of the Korea Research Foundation 1993 and the Basic Science Research Institute Program, Korean Ministry of Education 1994, Project No. BSRI-94-5408. TWJ and AF were supported in part by the NSF through grants AST91-00486 and AST93-18959, by NASA through grant NAGW-2548 and by the University of Minnesota Supercomputer Institute.

– 17 –## REFERENCES

Bell, J. B., Colella, P., & Tragenstein, J. A., 1989, J. of Comput. Phys., 82, 362.

Bracewell, R. N., 1986, The Fourier Transform and Its Application, (New York: McGraw-Hill).

Brackbill J. U., & Barnes, D. C., 1980, J. Comp. Phys., 35, 426.

Brio, M., & Wu, C. C., 1988, J. Comp. Phys., 75, 400.

Chevalier, R. A., Blondin, J. M. & Emmering, R. T., 1992, ApJ, 392, 118.

Colella, P., & Woodward, P. R., 1984, J. Comp. Phys., 54, 174.

Dahlburg, R. B., & Picone, J. M., 1989, Phys. of Fluids B, 1, 2153.

Dai W., & Woodward P. R., 1994a, J. of Comput. Phys., 111, 354.

Dai W., & Woodward P. R., 1994b, preprint.

DeVore, C. R., 1991, J. of Comput. Phys., 92, 142.

Frank, A., Jones, T. W., & Ryu, D., 1994, ApJS, 90, 975.

Frank, A., Jones, T. W., & Ryu, D., 1995a, ApJ, 441, 629.

Frank, A., Jones, T. W., & Ryu, D., 1995b, in preparation.

Godunov, S. K., 1959, Math. Sb., 47, 271.

Harten, A., 1983, J. Comp. Phys., 49, 357.

Harten, A., Engquist, B., Osher, S., Chakravarthy, S. R., 1987, J. Comp. Phys., 71, 231.

Ishii, T., Matsuda, T., Shima, E., Livio, M., Anzer, U., & Börner, G., 1993, ApJ, 404, 706.

Jackson, 1975, Classical Electrodynamics, (New York: John Wiley & Sons, Inc.).

Jeffrey A., & Taniuti, T., 1964, Nonlinear Waves Propagation, (New York: Academic Press).

Lind, K. R., Payne, D. G., & Meier, D. L., 1991, preprint.

Lind, K. R., Payne, D. G., Meier, D. L., & Blandford, R. D., 1989, ApJ, 344, 89.

Orszag, S. A., & Tang, C. M., 1979, J. Fluid Mech., 90, 129.

Miura, A. & Pritchett, P. L., 1982, JGR, 87, 7431.

Miura, A., 1984, JGR, 89, 801.

Picone, J. M., & Dahlburg, R. B., 1991, Phys. of Fluids B, 3, 29.

Porter, D. H., & Woodward, P. R., 1994, ApJS, 93, 309.

Roe, P. L., 1981, J. Comp. Phys., 43, 357.

Ryu, D., & Goodman, J., 1994, ApJ, 422, 269.

Ryu, D., & Jones, T. W., 1995, ApJ, 442, 228.

Ryu, D., Ostriker, J. P., Kang, H., & Cen R., 1993, ApJ, 414, 1.

– 18 –

## FIGURE CAPTIONS

**Fig. 1** The 2-dimensional MHD shock tube test. Structures propagate diagonally along the line $(0,0)$ to $(1,1)$ in the $x$-$y$ plane. The initial left state is $(\rho, v_\parallel, v_\perp, v_z, B_\perp, B_z, E) = (1, 10, 0, 0, 5/\sqrt{4\pi}, 0, 20)$ and the initial right state is $(1, -10, 0, 0, 5/\sqrt{4\pi}, 0, 1)$, with $B_\parallel = 5/\sqrt{4\pi}$ and $\gamma = 5/3$ (same test as Fig. 1a in Ryu & Jones 1995). Dots are from a numerical calculation with the MHD TVD code described in §2 using $256 \times 256$ cells and a Courant constant of 0.6. The structure is shown at time $t = 0.08\sqrt{2}$ along the grid diagonal. They are plotted along the diagonal line $(0,0)$ to $(1,1)$. Lines are from the nonlinear Riemann solver described in Ryu & Jones (1995). Plots show from left to right (1) fast shock, (2) slow rarefaction, (3) contact discontinuity, (4) slow shock, and (5) fast shock.

**Fig. 2** The $2+1/2$-dimensional MHD shock tube test. Structures propagate diagonally along the line $(0,0)$ to $(1,1)$ in the $x$-$y$ plane. The initial left state is $(\rho, v_\parallel, v_\perp, v_z, B_\perp, B_z, E) = (1.08, 1.2, 0.01, 0.5, 3.6/\sqrt{4\pi}, 2/\sqrt{4\pi}, 0.95)$ and the initial right state is $(1, 0, 0, 0, 4/\sqrt{4\pi}, 2/\sqrt{4\pi}, 1)$, with $B_\parallel = 2/\sqrt{4\pi}$ and $\gamma = 5/3$ (same test as Fig. 2a in Ryu & Jones 1995). Dots are from a numerical calculation with the MHD TVD code described in §2 using $256 \times 256$ cells and a Courant constant of 0.6. Structure is shown at time $t = 0.2\sqrt{2}$, along the diagonal line $(0,0)$ to $(1,1)$. Lines are from the nonlinear Riemann solver described in Ryu & Jones (1995). Plots show from left to right: (1) fast shock, (2) rotational discontinuity, (3) slow shock, (4) contact discontinuity, (5) slow shock, (6) rotational discontinuity, and (7) fast shock.

**Fig. 3** The time evolution of the spatially root-mean-squared $z$-magnetic field, $<\delta B_z^2>^{1/2}$, and $z$-velocity, $<\delta v_z^2>^{1/2}$, in the test of the decay of a two-dimensional Alfvén wave. An initial Alfvén wave has been set up in a stationary background and the decay has been followed in a numerical calculation using $32 \times 32$ cells. See text for details.

**Fig. 4** The normalized decay rates, $\Gamma_a L/c_a$, $\Gamma_f L/c_f$, and $\Gamma_s L/c_s$, and the Reynolds numbers, $R$ (see text for definition), versus resolution (the number of cells spanned by the box size $L$) in the tests of the decay of two-dimensional Alfvén, fast, and slow waves. Standing waves have been set up in a stationary background and the decay rates have been measured during their oscillation. The calculations have been done using $8 \times 8$, $16 \times 16$, $32 \times 32$, $64 \times 64$, $128 \times 128$, and $256 \times 256$ cells and a Courant constant of 0.6. See text for details. For comparison, dotted lines with $(\Gamma L/c_s) \propto n_{\rm cell}^{-2}$ and $R \propto n_{\rm cell}^{2}$ are drawn.

**Fig. 5** The normalized decay rate, $\Gamma L/c_f$, and the Reynolds number, $R$, versus the plasma $\beta$ in the test of the decay of a two-dimensional fast wave. A standing wave has been set up in a stationary background and the decay rate has been measured during its oscillation. The calculations have been done with the MHD TVD code for $B_o = 10, 3, 1, 0.3, 0.1, 0.03, 0.01$ 0.003, and 0.001. $128 \times 128$ cells has been used with a Courant constant of 0.6. For $B_o = 0$, the calculation has been done with the hydrodynamic TVD code and its result is indicated with dotted lines. See text for details.

**Fig. 6** The evolution of the spatially root-mean-squared transverse velocity $<v_y^2>^{1/2}$ in the MHD Kelvin Helmholtz instability. The results of simulations at three different resolutions are shown: (dash-dot = $128 \times 128$); (dashed = $256 \times 256$); (solid = $512 \times 512$). Initial conditions, which involve eigenmode perturbations on a smoothly sheared flow of uniform density and pressure are described in the text. Time is expressed in units of the predicted linear growth time, $\Gamma^{-1}$. The dotted line represents the predicted linear growth of the instability. The transverse velocity is normalized to the initial perturbation.

**Fig. 7** Grey scale Images of the MHD Kelvin-Helmholtz instability. These images were taken from the high resolution ($512 \times 512$) simulation at $t = 1.4/\Gamma$. Quantities shown (from left to right, top to bottom): gas density; magnetic pressure; gas pressure; magnetic field lines (contours of vector potential $A_z$).

**Fig. 8** Gray scale images of gas pressure (upper left), magnetic pressure (upper right), $\nabla \cdot \boldsymbol{v}$ (lower left), and $(\nabla \times \boldsymbol{v})_z$ (lower right) in the compressible Orszag-Tang vortex test. White indicates the regions with high (or positive) values and black indicates the regions with low (or negative) values. The calculation has been done in a periodic box with $x = [0, 1]$ and $y = [0, 1]$ using $256 \times 256$ cells and a Courant constant of 0.6. The initial configuration has been set with $\rho = 25/36\pi$, $p = 5/12\pi$, $\boldsymbol{v} = -\sin(2\pi y)\hat{\boldsymbol{x}} + \sin(2\pi x)\hat{\boldsymbol{y}}$, $\boldsymbol{B} = [-\sin(2\pi y)\hat{\boldsymbol{x}} + \sin(4\pi x)\hat{\boldsymbol{y}}]/\sqrt{4\pi}$, and $\gamma = 5/3$ and the plot shown is at $t = 0.48$.